\documentclass[aps,prb,twocolumn,showpacs,showkeys,preprintnumbers,groupedaddress,amsmath,amssymb,]{revtex4-1}

\usepackage{graphicx}
\usepackage{dcolumn}
\usepackage{bm}
\usepackage[dvips]{color}

\begin{document}




\title{Observation of double percolation transitions in Ag-SnO$_2$ nanogranular films}


\author{Yan-Fang Wei}
\author{Zhi-Qing Li}
\email[Author to whom correspondence should be addressed. Electronic address: ]{zhiqingli@tju.edu.cn}
\affiliation{Tianjin Key Laboratory of Low Dimensional Materials Physics and
Preparing Technology, Department of Physics, Tianjin University, Tianjin 300072,
China}


\date{\today}
\begin{abstract}
Two percolation transitions are observed in Ag$_x$(SnO$_2)_{1-x}$ nanogranular films with Ag volume fraction $x$ ranging from $\sim$$0.2$ to  $\sim$$0.9$. In the vicinity of each percolation threshold $x_{ci}$ ($i$$=$1, 2), the variation in $\sigma$ with $x$ obeys a power law for $x$$>$$x_{ci}$. The origin of the first percolation transition at $x_{c1}$ ($x_{c1}$$>$$x_{c2}$) is similar to that of the classical one, while the second transition is explained as originating from the tunneling to the second-nearest neighboring Ag particles. These observations provide strong experimental support for the validity of current theories concerning tunneling effect in conductor-insulator nanogranular composites.
\end{abstract}
\pacs{64.60.ah, 73.40.Gk, 72.80.Tm, 81.05.Rm}
\keywords{Percolation, Granular film, Tunneling effect, Composite materials}

\maketitle

Granular thin films, consisting of immiscible metal particles and insulating medium (usually amorphous), are a class of functional materials, and have attracted renewed attention during the last decades. The ease of adjusting metal grain size and the ratio of the metal to insulator make their electrical, magnetic, optical properties be tuned facilely. Hence they are not only potential materials applied in spintronic nanodevices,\cite{spintronic nanodevices} biosensors,\cite{biosensors} electrically resetable switches,\cite{electrically resetable switches1,electrically resetable switches2} recording media,\cite{recording media} catalysis,\cite{catalysis} and antistatic coatings\cite{antistatic coatings} but also suitable systems in studying mesoscopic physics.
In fundamental research side, it has been realized that the granular composites can reveal a series of unusual phenomena, such as giant Hall effect,\cite{giant Hall effect1,giant Hall effect2,giant Hall effect3} giant magnetoresistance effect,\cite{giant magnetoresistance1,giant magnetoresistance2} and superparamagnetism,\cite{superparamagnetism} and then produce new physics that are absent in homogeneous materials.
\par
Recently, great attention has been focussed on the percolation behavior of granular systems with nanoscale grain size.\cite{tunneling1,tunneling2,tunneling3,tunneling4,tunneling5,tunneling6,tunneling7}
The classical percolation theory on the transport properties of granular systems is based on the assumption that the electrons can transmit from one particle to the others only if they are geometrically connected. Then there exists a specific metal volume fraction $x_c$ (the percolation threshold) below which the conductivity is zero while above which the conductivity obeys $\sigma$$\simeq$$\sigma_{0}(x$$-$$x_c$)$^{t}$, where $\sigma_0$ is a proportionality constant, $t$$\simeq$$2$ is a critical exponent.\cite{classical theo} Though this scenario apparently neglects the tunneling effect among the metal particles, the power law of the conductivity has often been observed experimentally even if the tunneling effect was involved.\cite{tunneling1,tunneling2,tunneling3} Trying to understand the influence of tunneling effect on percolation system, Balberg, Grimaldi, and co-workers have intensively studied the percolation behavior of conductor-insulator composites in the framework of global tunneling network (GTN) model, in which each conducting particle can be connected to all the others via tunneling processes.\cite{tunneling4,tunneling5,tunneling6,tunneling7} Their results indicate that the behavior of the composites being percolationlike or tunnelinglike (hoppinglike) depends on the microstructure and the ratio of conducting particle size $D$ to the tunneling length $\xi$.\cite{tunneling4}  Particularly, when the microstructure of the composite systems is similar to that of the lattice model\cite{tunneling4,lattice model} and $D$ is several times (but greater than five) or tens times larger than $\xi$, multiple percolation thresholds would appear.\cite{tunneling6}  Experimentally, only the Ni-SiO$_{2}$ granular system has been reported to exhibit the trace of multiple percolation thresholds.\cite{tunneling4,tunneling5}  Hence the validity of the predication for multiple percolation  transitions still needs to be tested, and the microstructure and electrical properties of such granular systems deserve to be investigated systematically.
\par
In this letter, we report our experimental results of conductivity as functions of temperature and Ag volume fraction $x$ in Ag$_x$(SnO$_2$)$_{1-x}$ (0.2$\lesssim$$x$$\lesssim$0.9) nanogranular films. (Ag is immiscible with SnO$_2$, and Ag-SnO$_2$ composites are  potential contact materials used in relay.\cite{contact matereials1,contact matereials2}) It is found that two percolation thresholds appear in sequence with decreasing $x$. The microstructure as well as the transport mechanism at different $x$ regions  is also discussed.
\par
Ag-SnO$_2$ granular films were deposited on glass substrates at room temperature by a cosputtering method. An Ag and a SnO$_2$ targets
both of 60 mm in diameter were used as the sputtering sources. Details of the deposition processes were reported elsewhere.\cite{giant Hall effect3} More than 40 samples with $0.2$$\lesssim$$x$$\lesssim$$0.9$ have been fabricated and characterized. The thicknesses of the films ($\sim$500 nm), were determined by a surface profiler (Dektak, 6 M). The Ag volume fraction $x$ in each sample was obtained from the energy-dispersive x-ray spectroscopy analysis. The microstructure of the films was characterized by transmission electron microscopy (TEM, Tecnai G2 F20). The resistivities were measured by a standard four-probe method in a physical property measurement system (PPMS-6000, Quantum Design).
\begin{figure}[htp]
\begin{center}
\includegraphics[scale=0.9]{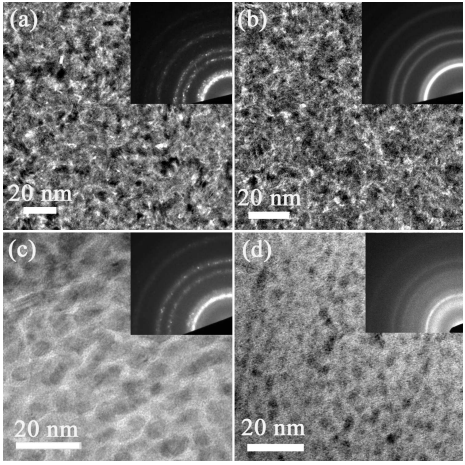}
\caption{Bright-field TEM images and SAED patterns
(insets) for Ag$_x$(SnO$_2$)$_{1-x}$ films with $x$ values of (a) 0.90, (b) 0.73, (c) 0.58, and (d) 0.26.}
\label{TEM}
\end{center}
\end{figure}
\par
Figure~\ref{TEM} shows bright-field TEM images and selected area electron diffraction (SAED)
patterns for four representative Ag$_x$(SnO$_2$)$_{1-x}$ films with $x$$=$0.90, 0.73, 0.58, and 0.26. For the films with large $x$ [Figs.~\ref{TEM}(a) and \ref{TEM}(b)], most of the Ag particles directly connect to others. While for the films with relatively small $x$ ($x$$\lesssim$$0.65$), Ag particles (the dark regions in the TEM images) are embedded in the SnO$_{2}$ matrix (bright regions), revealing typical granular characteristics [see Figs.~\ref{TEM}(c) and ~\ref{TEM}(d)].
The sharp and bright diffraction rings in the insets of Figs.~\ref{TEM}(a), \ref{TEM}(b), and \ref{TEM}(c) are corresponding to the diffraction of face-centered cubic Ag. The diffraction of SnO$_2$ is not present, indicating that  SnO$_2$ is amorphous. For films with $x$$\lesssim$$0.42$, Ag particles are also poorly crystallized [see the inset of Fig.~\ref{TEM}(d)]. The mean sizes of Ag grains for films with $0.5$$\lesssim$$x$$\lesssim$$0.65$ are $7$$\pm$$2$ nm, while they are  $5$$\pm$$2$ nm for the $x$$<$$0.42$ samples. We note in passing that a necessary condition for existing multiple percolation transitions in a granular system is that the sizes of metal
grains are limited in nanoscales.\cite{tunneling6}
\begin{figure}[htp]
\begin{center}
\includegraphics[scale=0.9]{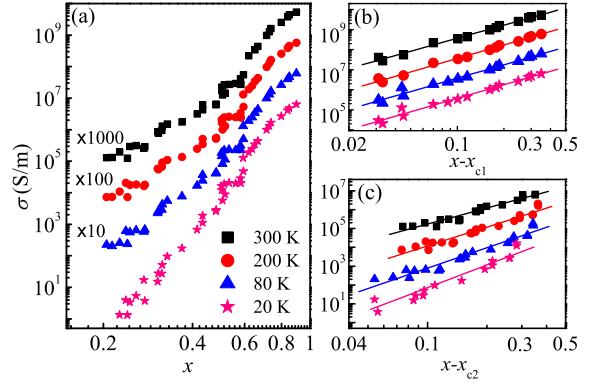}
\caption{(a) Conductivity vs metal volume fraction in double-logarithmic
scales at 300, 200, 80, and 20 K. (b) and (c) Log-log plots of
conductivity versus $x$$-$$x_{ci}$ [$i$$=$1 for (b) and $i$$=$2 for (c)]  at the same four temperatures as in (a). The
straight solid lines are least-squares fits to $\sigma$$=$$\sigma_{0i}$$(x$$-$$x_{ci})^{t_i}$.  For clarity, the data at 80, 200, and 300 K have been shifted up by multiplying a factor of 10, 100, and 1000, respectively.}
\label{r-x}
\end{center}
\end{figure}
\par
Figure~\ref{r-x}(a) shows the variation in conductivity $\sigma$ with $x$ in double-logarithmic scales at 20, 80, 200, and 300 K, as indicated. As $x$ increases from $\sim$$0.2$ to $\sim$$0.9$, the magnitude of $\sigma$ increases by a factor of nearly 7 (5) orders at 20 K (300 K).  At a given temperature, the $x$ dependence of $\sigma$ data can be divided into two distinct parts that are approximately in two different straight lines, and the boundary between the two parts is around $x$$=$$0.54$$\pm$$0.04$. The trend is more prominent at higher temperature. As mentioned above, the classical percolation theory predicates that $\sigma$ follows a power law with $x$, thus one can immediately obtain $\log$$\sigma$$=$$\log$$\sigma_{0}$$+$$t\log$$(x$$-$$x_c)$. Hence the fact that the Ag volume fraction $x$ dependence of $\sigma$ data are approximately in two distinct straight lines in axis with double-logarithmic scales strongly suggests that there are two percolation thresholds in our Ag-SnO$_2$ nanogranular films. Each part of the experimental $\sigma$ versus $x$ data is least-squares fitted to $\sigma$$\simeq$$\sigma_{0}(x$$-$$x_c$)$^{t}$, and the typical results are plotted in double-logarithmic scales in Figs.~\ref{r-x}(b) and \ref{r-x}(c), respectively. The fitting parameters $\sigma_{0i}$, $x_{ci}$, and $t_i$ obtained at different temperatures are listed in Table~\ref{Table}, where $i$$=$$1$ and $2$ refer to the fitted values at high and low $x$ parts, respectively. Inspection of Figs.~\ref{r-x}(b) and \ref{r-x}(c) indicates that both parts of the $\sigma$-$x$ data can be well described by the equation $\sigma$$\simeq$$\sigma_{0}(x$$-$$x_c$)$^{t}$. For the high $x$ part ($x$$\gtrsim$$0.55$), both the critical parameter $t_1$ and percolation threshold $x_{c1}$ are almost invariable with temperature from 300 down to 20 K. The critical exponent $t_1$$\simeq$$2.3$ is close to that deduced from the 3D classical percolation theory ($t^{th}$$\simeq$2), and $x_{c1}$$\simeq$$0.54$$\pm$$0.02$ is similar to that observed in other granular films.\cite{giant Hall effect2,tunneling4,tunneling5}  While for the low $x$ part ($0.2$$\lesssim$$x$$\lesssim$$0.5$), the critical parameter $t_2$ is much greater than the universal value 2 and the percolation threshold $x_{c2}$ increases with decreasing temperature (see further remarks below).
\begin{figure}[htp]
\begin{center}
\includegraphics[scale=0.9]{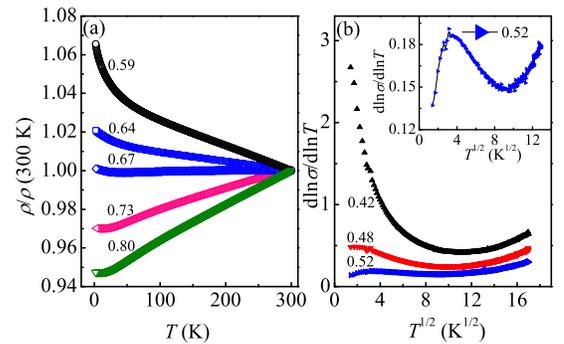}
\caption{(a) Normalized resistivity as a function of temperature for Ag$_x$(SnO$_2$)$_{1-x}$ films with $x$$=$0.59, 0.64, 0.67, 0.73, and 0.80.    (b) $d\ln \sigma/d\ln T$ versus $T^{1/2}$ for films with $x$$=$0.42, 0.48, and 0.52. Inset: an enlarged plot of $d\ln \sigma/d\ln T$ versus $T^{1/2}$ for the film with $x$$=$0.52.}
\label{r-t}
\end{center}
\end{figure}
\par
We detailedly discuss the transport processes related to the two percolation transitions now. Figure 3(a) shows the normalized resistivity as a function of temperature for five representative films with $x$$>$$x_{c1}$ ($x$$=$0.59, $0.64$, 0.67, 0.73, and 0.80), as indicated. For $x$$\gtrsim$0.67, the resistivity decreases with decreasing temperature, reaches its minimum and then slightly increases with further decreasing temperature, which is the typical behavior of a dirty metal. Thus one can readily conclude that the conducting paths are formed by the connected Ag particles for $x$$\gtrsim$$0.67$. For samples with $0.54$$\lesssim$$x$$\lesssim$$0.65$, the temperature coefficients of resistivities are negative at the whole measuring temperature range, indicating that most of the direct conducting paths are broken down and the tunneling effect governs the electrical transport processes at this regime.\cite{tunneling broken} The enhancement in the magnitude of resistivity with decreasing temperature may relate to the interparticle electron-electron Coulomb interaction.\cite{coulomb effect1,coulomb effect2,e-e exp1,e-e exp2} Considering the logarithmic derivative $w$$=$$d\ln \sigma/d\ln T$ of the conductivity is more sensitive than the temperature coefficient of conductivity $d\sigma/dT$ and defines a more accurate and reliable criterion to distinguish between metallic and insulating behavior,\cite{w} we got the logarithmic derivative $w$ of the films and presented the representative ones in Fig.~\ref{r-t}(b), as indicated. Clearly, the magnitudes of $w$ tend to vanish, be a constant, and be divergent for the $x$$=$$0.52$, $0.48$, and 0.42 films, respectively, as $T$ approaches to zero, indicating that films with $x$$\lesssim$$0.48$ and $x$$\gtrsim$$0.52$ are insulators and metals in electrical transport properties, respectively. Hence the metal-insulator transition in Ag$_x$(SnO$_2$)$_{1-x}$ granular system occurs at $x_{\text{M-I}}$$\simeq$$0.5$, which is slightly less than (but close to) the value of $x_{c1}$ ($\simeq$$0.54$$\pm$0.02).  The comparability of $x_{\text{M-I}}$ and $x_{c1}$ confirms the reliability of the percolation threshold value $x_{c1}$ obtained from fits to the power law.
\begin{table}[htp]
\caption{Some fitting parameters for Ag$_x$(SnO$_2$)$_{1-x}$ nanogranular films. Here $\sigma_{0i}$, $x_{ci}$, and $t_{i}$ ($i$$=$$1,2$) are defined in the text, $D/\xi$ is the adjusting
parameter in Eq.~(\ref{Eq.(Hopping)}). The uncertainties are $\approx$$2\%$, 6\%, 20\%, and 8\% in $x_{c1}$, $t_1$, $x_{c2}$, and $t_2$, respectively.}
\label{Table}
\begin{ruledtabular}
\begin{center}
\begin{tabular}{ccccccccc}
$T$&   $\sigma_{01}$& $x_{c1}$& $t_{1}$& $\sigma_{02}$& $x_{c2}$& $t_{2}$&   $D/\xi$\\
(K)&   ($10^7$ S/m)&   &        &       ($10^5$ S/m)&          &        &      \\\hline
300&   5.8&           0.55&	      2.2&	  1.7&	         0.13&	      3.4&    7.5\\
200&   6.6&           0.55&	      2.3&	  2.0&	         0.14&	      3.6&    8.2\\
150&   7.0&           0.54&	      2.4&	  2.1&	         0.14&	      3.6&    8.7\\
80&	   7.6&           0.54&	      2.4&	  2.9&	         0.16&	      3.7&   10.1\\
50&    7.9&           0.54&	      2.4&	  3.9&	         0.18&	      3.7&   11.5\\
30&    8.0&           0.55&	      2.4&	  5.3&	         0.21&	      4.1&   13.5\\
20&	   8.1&           0.54&	      2.4&	  9.5&	         0.22&	      3.9&   15.4\\
\end{tabular}
\end{center}
\end{ruledtabular}
\end{table}
\par
To solve the percolation and tunneling issues in conductor-insulator composites, Ambrosetti \emph{et al.}\cite{tunneling6} considered two kinds of ideal microstructure within the GTN consideration: the lattice model and the continuum model.\cite{tunneling5,tunneling6,continuum model} In the lattice case, the conductivity of the metal-insulator granular composites with large  conducting particle size ($D/\xi$$\gtrsim$100) would drop dramatically and eventually tend to zero with decreasing metal volume fraction $x$ below the percolation threshold $x_{c}$, while it follows a $\sigma_0 (x$$-$$x_c)^{t}$ law in the $x$$>$$x_c$ region. This is the often observed case in metal-insulator granular systems.  However, for composites with nanoscale conducting particles, the tunneling to the second-nearest neighboring particles cannot be ignored, then the reduction of conductivity with decreasing $x$ would be punctuated by two remarkable drops at $x$$=$$x_{ci}$ ($i$$=$$1,2$). In the vicinity of $x_{ci}$, the conductivity obeys the power law, $\sigma$$=$$\sigma_{0i}$$(x$$-$$x_{ci})^{t_i}$, for $x$$>$$x_{ci}$. This is the so-called multiple percolation thresholds phenomenon. If the tunneling to the $n$th-nearest neighboring particles has to be considered, the number of  percolation thresholds would be $n$. For our Ag$_x$(SnO$_{2}$)$_{1-x}$ granular films with $x$$<$$0.42$, the mean size of Ag particles is $5$$\pm$$2$ nm and the tunneling length $\xi$ is less than one nanometer, hence the tunneling effect to the third-nearest neighboring particles can be safely ignored and only two percolation thresholds may be observed. Combining the experimental observations mentioned above, one can see that the lattice model is quite applicable to our Ag$_x$(SnO$_{2}$)$_{1-x}$ granular films though it only represents an ideal case of the true system.  On the other hand, when the microstructure of the nanoscale composites is similar to that of the continuum model, the conductivity would decrease smoothly with decreasing $x$ and not obey power law anymore. For composites with sufficiently small $x$ values, the variation in $\sigma$ with $x$ can be written as,
\cite{tunneling6,tunneling7}
\begin{equation}\label{Eq.(Hopping)}
\sigma(x)=\sigma_{0}'\exp\left(-1.41\frac{D}{\xi x^{1/3}}\right),
\end{equation}
where $\sigma_{0}'$ is a prefactor independent of temperature. For the nanoscale (but $D/\xi$$>$5) conductor-insulator composites, although the conductivity versus $x$ curve  deduced from the lattice model does not overlap with that from the continuum model [Eq.~(\ref{Eq.(Hopping)})], it  disperses in a zigzag course around the curve given by Eq.~(\ref{Eq.(Hopping)}).
\begin{figure}[htp]
\begin{center}
\includegraphics[scale=0.8]{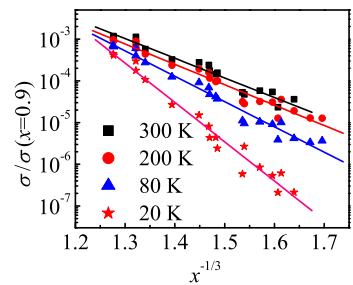}
\caption{Normalized conductivity versus $x^{-1/3}$ at different temperatures for Ag$_x$(SnO$_{2}$)$_{1-x}$ films ($0.21$$\lesssim$$x$$\lesssim$$0.48$). The straight solid lines are least-squares fits to  Eq.~(\ref{Eq.(Hopping)}).}
\label{FIGRx}
\end{center}
\end{figure}
\par
Figure~\ref{FIGRx} shows the logarithm of the normalized conductivity
as a function of $x^{-1/3}$ at 300, 200, 80, 20 K for Ag$_x$(SnO$_{2}$)$_{1-x}$ films with $0.21$$\lesssim$$x$$\lesssim$$0.48$. The straight solid  lines are the fitting results of Eq.~(\ref{Eq.(Hopping)}). Clearly, the experimental data at each temperature are distributed beside the straight line in a zigzag course. This further validates the lattice model in our Ag$_x$(SnO$_{2}$)$_{1-x}$ granular films. The fitting parameter $D/\xi$ is also listed in Table~\ref{Table}. The value of $D/\xi$ decreases with increasing temperature. Particularly, when temperature increases from 20 to 300 K, the value of $D/\xi$  decreases from $\sim$15 to $\sim$7.5. According to Ambrosetti \emph{et al.},\cite{tunneling6} this is exactly in the proper $D/\xi$ value range to appear multiple percolation transitions.
Since the mean size of Ag grains $D$ is insensitive to temperature between 20 to 300 K, the above observation also indicates that the tunneling length $\xi$ increases with increasing temperature. As a result, the value of the second percolation threshold $x_{c2}$ increases with decreasing temperature. The exponent $t_2$ ($\simeq$$3.7$$\pm$$0.4$) is much greater than the universal value $2$, which can be obtained within the GTN model by allowing a finite dispersion in the distances of the second-nearest neighboring metal particles.\cite{tunneling9,tunneling10}
\par
In summary, we systematically investigated the metal volume fraction $x$ and temperature dependences of conductivities in a series of Ag$_x$(SnO$_{2}$)$_{1-x}$ nanogranular films. It is found that two percolation thresholds, $x_{c1}$ and $x_{c2}$, are sequently present with decreasing $x$. In the vicinity of each percolation threshold $x_{ci}$, the variation in $\sigma$ with $x$ obeys $\sigma$$=$$\sigma_{0i}$$(x$$-$$x_{ci})^{t_i}$ for $x$$>$$x_{ci}$. Comparing with the current tunneling and percolation theories, we have found that the origin of the first percolation transition is similar to that of the classical percolation transition, while the second one is due to the tunneling to the second-nearest neighboring Ag particles. One reason for revealing two percolation transitions phenomenon in our Ag-SnO$_{2}$ nanogranular films is the comparability between the microstructure of the films and that of ideal lattice model, the other is the ideal ratio of mean size of Ag particles to the tunneling length in SnO$_2$ matrix.
\par
The authors are grateful to J. J. Lin for valuable
discussions. This work was supported by the NSF of China through grant no. 11174216 and Research Fund for the Doctoral Program of Higher Education.


\begin{thebibliography}{00}\label{sec:TeXbooks}
\bibitem{spintronic nanodevices} S. Mitani, S. Takahashi, K. Takahnashi, K. Yakushiji, S. Maekawa, and H. Fujimori, Phys. Rev. Lett. \textbf{81}, 2799 (1998).
\bibitem{biosensors} M. Huth, J. Appl. Phys. \textbf{107}, 113709 (2012).
\bibitem{electrically resetable switches1} I. Balberg, D. Azulay, D. Toker, and O. Millo, Int. J. Mod. Phys. B \textbf{18}, 2091 (2004).
\bibitem{electrically resetable switches2} D. Azulay, M. Eylon, O. Eshkenazi, D. Toker, M. Balberg, N. Shimoni, O. Millo, and I. Balberg, Phys. Rev. Lett. \textbf{90}, 236601 (2003).
\bibitem{recording media} S. Stavroyiannis, I. Panagiotopoulos, D. Niarchos, J. A. Christodoulides, Y. Zhang, and G. C. Hadjipanayis, Appl. Phys. Lett. \textbf{73}, 3453 (1998).
\bibitem{catalysis} A. H. Lu, E. L. Salabas, and F. Sch\"{u}th, Angew. Chem. Int. Edn. \textbf{46}, 1222 (2007).
\bibitem{antistatic coatings} R. Wycisk, R. Po\'{z}niak, and A. Pasternak, J.  Electrost. \textbf{56}, 55 (2002).
\bibitem{giant Hall effect1} A. B. Pakhomov, X. Yan, and B. Zhao, Appl. Phys. Lett. \textbf{67}, 3497 (1995).
\bibitem{giant Hall effect2} X. X. Zhang, C. Wan, H. Liu, Z. Q. Li, P. Sheng, and J. J. Lin, Phys. Rev. Lett. \textbf{86}, 5562 (2001).
\bibitem{giant Hall effect3} Y. N. Wu, Z. Q. Li, and J. J. Lin, Phys. Rev. B \textbf{82}, 092202 (2010).
\bibitem{giant magnetoresistance1} A. E. Berkowitz, J. R. Mitchell, M. J. Carey, A. P. Young, S. Zhang, F. E. Spada, F. T. Parker, A. Hutten, and G. Thomas, Phys. Rev. Lett. \textbf{68}, 3745 (1992).
\bibitem{giant magnetoresistance2} J. Q. Xiao, J. S. Jiang, and C. L. Chien, Phys. Rev. Lett. \textbf{68}, 3749 (1992).
\bibitem{superparamagnetism} D. H. Ping, M. Ohnuma, K. Hono, M. Watanabe, T. Iwasa, and T. Masumoto, J. Appl. Phys. \textbf{90}, 4708 (2001).
\bibitem{classical theo} \emph{Percolation Structures and Processes}, edited by G. Deutscher, R. Zallen, and J. Adler (Adam Hilger, Bristol, 1983), p. 207.
\bibitem{tunneling1} J. M. Normand and H. J. Herrmann, Int. J. Mod. Phys. C \textbf{6}, 813 (1995).
\bibitem{tunneling2} S. V. Menot, C. Grimaldi, T. Meader, S. Str\"{a}ssler, and P. Ryser, Phys. Rev. B \textbf{71}, 064201 (2005).
\bibitem{tunneling3} P. Sheng, E. K. Sichel, and J. I. Gittleman, Phys. Rev. Lett. \textbf{40}, 1197 (1978).
\bibitem{tunneling4} D. Toker, D. Azulay, N. Shimoni, I. Balberg, and O. Millo, Phys. Rev. B \textbf{68}, 041403 (2003).
\bibitem{tunneling5} I. Balberg, J. Phys. D: Appl. Phys. \textbf{42}, 064003 (2009).
\bibitem{tunneling6} G. Ambrosetti, I. Balberg, and C. Grimaldi, Phys. Rev. B \textbf{82}, 134201 (2010).
\bibitem{tunneling7} G. Ambrosetti, C. Grimaldi, I. Balberg, T. Maeder, A. Danani, and P. Ryser, Phys. Rev. B \textbf{81}, 155434 (2010).
\bibitem{lattice model} In lattice model, the conducting particles randomly occupy a fraction of the sites of a regular lattice. If two nearest nighbour sites are both occupied, they are called electrically connected.
\bibitem{contact matereials1} L. Chaffron, Y. Chen, and G. Martin, Ann. Chim. \textbf{18}, 395 (1993).
\bibitem{contact matereials2} A. Verma, A. Roy, and T. R. Anantharaman, Int. J. Powder Metall. \textbf{27}, 51 (1991).
\bibitem{tunneling broken} S. Mitani, H. Fujimori, and S. Ohnuma, J. Magn. Magn. Mater. \textbf{165}, 141 (1997).
\bibitem{coulomb effect1} K. B. Efetov and A. Tschersich, Phys. Rev. B \textbf{67}, 174205 (2003); Europhys. Lett. \textbf{59}, 114 (2002).
\bibitem{coulomb effect2} I. S. Beloborodov, A. V. Lopatin, V. M. Vinokur, and K. B. Efetov, Rev. Mod. Phys. \textbf{79}, 469 (2007).
\bibitem{e-e exp1} Y. J. Zhang, Z. Q. Li, and J. J. Lin, Phys. Rev. B \textbf{84}, 052202 (2011).
\bibitem{e-e exp2} Y. Yang, X. D. Liu, and Z. Q. Li, Appl. Phys. Lett. \textbf{100}, 262101 (2012).
\bibitem{w} A. M\"{o}bius, Phys. Rev. B \textbf{40}, 4194 (1989).
\bibitem{continuum model} In continuum model, the conducting particles are dispersed with an equilibrium distribution in a continuous insulator medium.
\bibitem{tunneling9} N. Johner, C. Grimaldi, I. Balberg, and P. Ryser, Phys. Rev. B \textbf{77}, 174204 (2008).
\bibitem{tunneling10} C. Grimaldi and I. Balberg, Phys. Rev. Lett. \textbf{96}, 066602 (2006).
\end{thebibliography}
\end{document}